\documentclass[aps,superscriptaddress,twocolumn,nofootinbib]{revtex4}
\usepackage{graphicx}

\makeatletter

\begin{document}

\title{Economics: the next physical science?}
\author{J. Doyne Farmer}
\affiliation{Santa Fe Institute, 1399 Hyde Park Rd., Santa Fe NM 87501}  
\author{Martin Shubik}
\affiliation{Economics Department, Yale University, New Haven CT}
\author{Eric Smith}
\affiliation{Santa Fe Institute, 1399 Hyde Park Rd., Santa Fe NM 87501}
\date{\today}

\begin{abstract}
We review an emerging body of work by physicists addressing questions of
economic organization and function. We suggest that, beyond simply employing
models familiar from physics to economic observables, remarkable
regularities in economic data may suggest parts of social order that can
usefully be incorporated into, and in turn can broaden, the conceptual
structure of physics.
\end{abstract}

\maketitle
\tableofcontents

\affiliation{Santa Fe Institute, 1399 Hyde Park Rd., Santa Fe NM 87501}

\affiliation{Economics Department, Yale University, New Haven CT}

\affiliation{Santa Fe Institute, 1399 Hyde Park Rd., Santa Fe NM 87501}

\section{Physics and economics}

In the last decade or so physicists have begun to do academic research in
economics in a newly emerging field often called ``econophysics''. Perhaps a
hundred people are now actively involved, with two new journals\footnote{%
The new journals are The International Journal of  Pure and Applied Finance
and Quantitative Finance. Quantitative  Finance was started by the Institute
of Physics and later sold to a  commercial firm.} and frequent conferences.
At least ten books have been written on econophysics in general or specific
subtopics (We are restricted by citation limits here, but an extensive
bibliography of the books and archived articles is maintained on the
econophysics website, http://www.unifr.ch/econophysics). Ph.D.~theses are
being granted by physics departments for research in economics, and in
Europe there are several professors in physics departments specializing in
econophysics. There is even a new annual research prize, titled the ``Young
Scientist Award for Social and Econophysics''. Is this just a fad, or is
there something more substantial here?

If physicists want to do research in economics, why don't they just get
degrees in economics in the first place? Why don't the econophysicists
retool, find jobs in economics departments and publish in traditional
economics journals? Perhaps this is just a temporary phenomenon, driven by a
generation of physicists who made a bad career choice. Is there any reason
why research in economics should be done in physics departments as an
on-going activity, or why economics departments should pay heed to the
methods of physics? What advantage, if any, is conferred by a background in
physics? And most important, how does econophysics differ from economics,
and what unique contribution can it make, if any?

One is naturally suspicious that the emergence of econophysics is just a
reflection of a depressed job market. It is certainly true that during the
last two decades a large number of physicists have been lured to Wall St,
and that this has been an important stimulus. But this is not the main
focus. Econophysics is primarily an academic endeavor, whose participants
are employed by universities. It offers no special advantages in the job
market -- in fact, quite the opposite: It is even more competitive than
mainstream fields of physics. No permanent positions in econophysics have
ever been offered in an American university. Papers submitted to Physical
Review Letters require special justification concerning their relevance to
physics. While the situation in Europe is a little better than in the U.~S.,
jobs are still very scarce. The tenuous existence of econophysics relies on
senior professors who have redirected their interests from other areas, as
well as a few bold students and postdocs.

The involvement of physicists in social science has a long history, going
back at least to Daniel Bernoulli, who in 1738 introduced the idea of
utility to describe people's preferences. In his \textit{Essai philosophique
sur les probabilites} (1812), Laplace pointed out that events that might
seem random and unpredictable, such as the number of letters that end up in
the Paris dead-letter office, can in fact be quite predictable and can be
shown to obey simple laws. These ideas were further amplified by Quetelet, a
student of Fourier, who in 1835 coined the term ``social physics'', and
studied the existence of patterns in data sets ranging from the frequency of
different methods for committing murder to the chest size of Scottish men.
Analogies to physics played an important role in the development of economic
theory through the nineteenth century, and some of the founders of
neoclassical economic theory\footnote{%
``Neoclassical economics'' refers to a representation of individual decision
making in terms of scalar ``utility'' functions, whose gradients are
imagined to be like forces directing people to trade, and from which
economic equilibria arise as a kind of ``force balance'' among different
people's trading wishes. From the economist's point of view neoclassical
economics clarified and extended the work of the classical economists,
Smith, Mill, Ricardo and others by formalizing the notions of competition,
marginal utility and rent, as well as producing separate theories of the
firm and consumer.'' }, such as Irving Fisher, a student of Willard Gibbs, were originally
trained as physicists. Ettore Majorana in 1938 presciently outlined both the
opportunities and pitfalls in applying statistical physics methods to the
social sciences.

The range of topics that have been addressed spans many different areas of
economics. Finance is particularly well represented; sample topics include
the empirical observation of regularities in market data, the dynamics of
price formation, the understanding of bubbles and panics, methods for
pricing derivatives such as options, the construction of optimal portfolios,
and many other subjects. Broader topics in economics include the
distribution of income, theories of how money emerges, and implications of
symmetry and scaling to the functioning of markets.

Despite their long history of association, we see the substantial
contributions of physics to economics as still in an early stage, and find
it fanciful to forecast what will ultimately be accomplished. Almost
certainly, ``physical'' aspects of theories of social order will \emph{not}
simply recapitulate existing theories in physics, though already there
appear to be overlaps. The development of societies and economies is
potentially contingent on accidents of history, and at every turn hinges on
complex aspects of human behavior. Nonetheless, striking empirical
regularities such as those we survey below suggest that at least some social
order is not historically contingent, and perhaps is predictable from first
principles. The role of markets as mediators of communication and
distributed computation, and the emergence of the social institutions that
support them, are quintessentially economic phenomena. Yet the notions of
their computational or communication capacity, and how these account for
their stability and historical succession, may naturally be parts of the
physical world as it includes human social dynamics. In the context of human
desires, markets and other economic institutions bring with them notions of
efficiency or optimality in satisfying those desires. While intuitively
appealing, such notions have proven hard to formalize, and the examples
below show some progress in this area. As with most new areas of physical
inquiry, we expect that the ultimate goals of a ``physical economics'' will
be declared with hindsight, from successes in identifying, measuring,
modeling and in some cases predicting empirical regularities.

\section{Data analysis and the search for empirical regularities}

Economists are typically better trained in statistical analysis than
physicists, so this might seem to be an area where physicists have little to
contribute. However, differences in goals and philosophy are important.
Physics is driven by the quest for universal laws. In part, because of the
extreme complexity of phenomena in society, in the postmodern world where
relativist philosophies of science enjoy disturbingly widespread acceptance,
this quest has been largely abandoned. Modern work in social science is
largely focused on documenting differences. Although this trend is much less
obvious in economics, a typical paper in financial economics, for example,
might study the difference between the NYSE and the NASDAQ stock exchanges,
or the effect of changing the tick size of prices (the unit of the smallest
possible price change). Physicists have (perhaps naively) entered with fresh
eyes and new hypotheses, and have looked at economic data with the goal of
finding pervasive regularities, emphasizing what might be common to all
markets rather than what might make them different. This work has been
opportunistically motivated by the the existence of large data sets such as
the complete transaction histories of major exchanges over timespans of
years, which in some cases contain hundreds of millions of events.

\begin{figure}[ht]
\begin{center}
\includegraphics[scale=0.4,angle=-90]{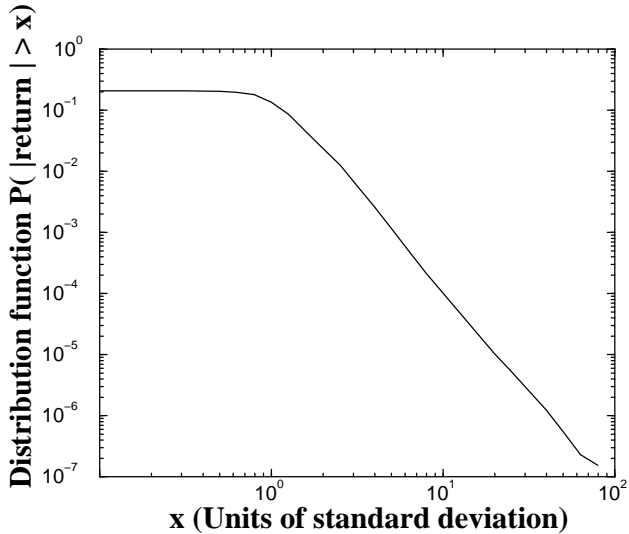}
\end{center}
\caption{ The distribution of 15 minute price movements for 1000 stocks from
the NYSE and NASDAQ, from ref.~\protect\cite{Gabaix03}. A movement in price $%
p$ is defined as $\log p(t + \tau) - \log p(t)$. The price movements for
each stock were normalized by dividing by the standard deviation for that
stock. There is a total of 15 million events. The straight line on the right
side of the curve indicates a power law, over about two orders of magnitude. 
}
\label{fig:power_law_fig}
\end{figure}

Much of the work by physicists in economics concerns power laws. A power law
is an asymptotic relation of the form $f(x)\sim x^{-\alpha }$, where $x$ is
a variable and $\alpha >0$ is a constant. In many important cases $f$ is a
probability distribution. Power laws have received considerable attention in
physics because they indicate scale free behavior and they are
characteristic of critical or nonequilibrium phenomena. In fact, the first
power law distribution (in any field) was observed in economics (see Box~\ref
{box:income_dist}), and the existence of power laws in economics has been a
matter of debate ever since. In 1963 Benoit Mandelbrot~\cite{Mandelbrot97}
observed that the distribution of cotton price fluctuations follows a power
law. Further observations of power laws in price changes were subsequently
noted by Rosario Mantegna and Eugene Stanley~\cite{Mantegna99}, who coined
the term ``econophysics''. Fig.~\ref{fig:power_law_fig} shows the striking
fidelity often found in economic power laws. The existence of power laws in
price changes is interesting from a practical point of view because of its
implications for the risk of financial investments, and from a theoretical
point of view because it suggests scale independence and possible analogies
to nonequilibrium behavior in the processes that generate financial returns. 

Since then many other power laws have been discovered by physicists. These
include the variance in the growth rates of companies as a function of their
size \cite{Amaral97}, the distribution function for the number of shares in
a transaction, the distribution for the number of trading orders submitted
at a price $x$ away from the best price offered, the size of the price
response to a trade as a function of the size $C$ of the company being
traded, and many other examples. The question of why power laws are so
ubiquitous in financial markets has stimulated a great deal of theoretical
work.

One of the most famous and most used models of prices is the random walk.
The random walk was originally introduced for prices in 1900 by Louis
Bachelier, a student of Poincar\'{e}, five years before Einstein introduced it
to describe Brownian motion. This forms the basis for the Black-Scholes
theory of option pricing~\cite{Bouchaud00}, which won a Nobel prize in
economics in 1997 (see Box~\ref{box:option_pricing}). One of the interesting
and surprising properties of the random walk of real prices is that its
diffusion rate is not constant. The size of price changes is strongly
positively autocorrelated in time, a phenomenon called \textit{clustered
volatility}\footnote{%
The autocorrelation function of a time series $x(t)$ is defined as $C(\tau)
= \langle \tilde{x}(t)\tilde{x}(t -\tau) \rangle$, where $\tilde{x} = (x -
\langle x \rangle)/\sigma_x$, $\langle ~ \rangle$ denotes a time average and 
$\sigma_x$ the standard deviation. ``Volatility'' is the term used  in
finance to refer to the variance or size of price shifts.}. The
autocorrelation of the size of price changes decays as a power law of the
form ${\tau}^{-\gamma}$. Since $0<\gamma <1$, this is a long-memory process.
Long-memory processes display anomalous diffusion and very slow convergence
of statistical averages.

Physicists have recently discovered that volatility is just one of several
long-memory-processes in markets. One of the most surprising concerns
fluctuations in supply and demand \cite{Bouchaud04}. If one assigns $+1$ to
an order to buy and $-1$ to an order to sell, the resulting series of
numbers has a positive autocorrelation function that decays as a power law
with an exponent $\gamma \approx 0.6$, which persists at statistically
significant levels across tens of thousands of orders, for periods of time
lasting for weeks. This implies that changes in supply and demand obey a
long-memory process. This has interesting implications.

One of the most fundamental principles in financial economics is called 
\textit{market efficiency}. This principle takes many forms: A market is
informationally efficient if prices reflect all available information; it is
arbitrage efficient if it is impossible for investors to make ``excess
profits'', and it is allocationally efficient if prices are set so that they
in some sense maximize everyone's welfare. One of the consequences of
informational efficiency is that prices should not be predictable. In
reality this is not a bad approximation; even the best trading strategies
exploit only very weak levels of predictability.

The coexistence of the long-memory of supply and demand with market
efficiency creates an interesting and as yet unresolved puzzle. Long-memory
processes are highly predictable using a simple linear algorithm. Since the
entrance of new buyers tends to drive the price up, and the entrance of new
sellers tends to drive it down, this naively suggests that price changes
should also be long-memory, which would violate market efficiency. To
prevent this from happening, the agents in the market must somehow
collectively adjust their behavior to offset this, for example by creating
an asymmetric response of prices, so that when there is an excess of new
buyers the price response to new buy orders is smaller than it is to new
sell orders. How this comes about, and why it comes about, remains a
mystery. This may be related to the cause of clustered volatility.

There is a great deal of other empirical work using methods and analogies
from physics that we do not have the space to describe in any detail. For
example, random matrix theory~\cite{Burda} (developed in nuclear physics)
and the use of ultrametric correlations have proved useful for understanding
the correlation between the movement of prices of different companies. An
analogy to the Omori law for seismic activity after major earthquakes has
proved to be useful for understanding the aftermath of large crashes in
stock markets, and other analogies from geophysics has led to a
controversial hypothesis about why markets crash \cite{Sornette02}. The
statistics of price movements have been noted to bear a striking resemblance
to those of turbulent fluids, which has led to what may now be the best
empirical models available for predicting clustered volatility. Such
examples speak to the universality of mathematics in its applications to the
world.

\section{Modeling the behavior of agents}

The most fundamental difference between a physical system and an economy is
that economies are inhabited by people, who have strategic interactions.
Because people think, plan, and make decisions based on their plans, they
are much more complicated to understand than atoms. This is a problem that
physics has never coped with, and it has caused the mathematical techniques
and modeling philosophy in economics to diverge from those in physics. While
this is clearly necessary, many physicists would argue that the gap is wider
than it should be.

The central approach to the problem of strategic interactions in
neoclassical economics is the theory of rational choice. The economists'
stylized version of individual rationality is to maximize some measure of
one's personal (usually material) welfare, having perfect knowledge of the
world and of other agents' goals and abilities, and the ability to perform
computations of any complexity. When agent A considers any strategy, agent B
knows that A is considering that strategy, and A knows that B knows that A
knows, and so on. This infinite regress appears very complicated. However, a
key simplifying result is that in any game there exists at least one Nash
equilibrium, which is a set of strategies with the property that each is the
optimal response to all of the others.

The Nash equilibrium is a fixed point in the space of strategies, which
circumvents the infinite regress problem by imposing self-consistency as a
defining criterion. Subject to several caveats, rational players who are not
cooperating with each other will choose a Nash strategy. This is the
operational meaning of rational choice. The assumption that decisions of
real human beings can be approximated in this way dominated economic
thinking about individual choices (called \emph{microeconomics}) from 1950
until the mid-1980s, though it is clearly implausible for all but the
simplest cognitive tasks. It also leaves unaddressed the problem of
aggregation of individual choices and the behavior of large populations
(called \emph{\ macroeconomics}).

\section{The quest for simple models of non-rational choice}

In the last twenty years economics has begun to challenge the assumptions of
rational choice and perfect markets by modeling imperfections such as
asymmetric information, incomplete market structure, and bounded
rationality. Several new schools of thought have emerged. The behavioral
economists attempt to take human psychology into account by studying
people's actual choices in idealized economic settings. Another school uses
idealizations of problem solving and learning ranging from standard
statistical methods to artificial intelligence to address the problem of
bounded rationality. Agent-based modeling makes computer simulations based
on idealizations of human behaviors and focuses on the complexity of
economic interactions. Yet another approach assumes that some agents (called
noise traders) have extremely limited reasoning capabilities while others
are perfectly rational. Physicists have joined with many economists in
seeking new theories of non-fully-rational choice, bringing new perspectives
to bear on the problem.

An early effort using both agent based modeling and artificial intelligence
is called the Santa Fe Stock Market. This grew out of a conference in 1986
organized by Ken Arrow, Phil Anderson and David Pines \cite{Anderson88} that
brought together physicists and economists. Presaging the modern move toward
behavioral economics, the physicists all expressed disbelief in theories of
rational choice and suggested that the economists should take human
psychology and learning more into account. The Santa Fe Stock Market was a
collaboration between economists, physicists and a computer scientist that
grew out of this conference. It replaced the rational agents in an idealized
market setting with an artificial intelligence model \cite{Arthur97}. It
showed that this leads to qualitative modifications of the statistics of
prices, such as fat tailed distributions of price change and clustered
volatility, and suggested that non-rational behavior plays an important part
in generating these phenomena.

The problem with this approach is that it is complicated, and while it
captures some qualitative features of markets, the path to more quantitative
theories is not clear. Agent based models tend to require ad hoc assumptions
that are difficult to validate. The hypothesis of rational choice, in
contrast, has the great virtue that it is parsimonious, making strong
predictions from simple hypotheses. From this point of view it is more like
theories in physics. This perspective has inspired the search for other
simple parsimonious alternatives. One such approach is often called \textit{%
zero intelligence}. This amounts to the assumption that agents behave more
or less randomly, subject to constraints such as their budget. Zero
intelligence models can be used to study the properties of market
institutions, and to determine which properties of a market depend on
intentionality and which don't. This provides a benchmark to avoid getting
lost in the large space of realistic human behaviors. Once a zero
intelligence model has be made, it can be modified by incorporating more
realistic assumptions, adding a little intelligence based on empirical
observations or models of learning. Where rational choice enters the
wilderness of bounded rationality from the top, zero intelligence enters it
from the bottom.

The zero intelligence approach can be traced back to the work of Herbert
Simon, a Nobel laureate in economics and pioneer in artificial intelligence.
Its main champions in recent years have been physicists, who have used
analogies to statistical mechanics to develop new models of markets. A good
example is the work of Per Bak, Maya Paczuski, and Martin Shubik (two
physicists and an economist), who studied the impact of random trading
orders on prices within an idealized model of price formation. They assumed
that traders simply place orders to buy or sell at random above or below the
prices of the most recent transactions. They then modify their orders from
time to time, moving them toward the middle until they generate a
transaction. The result is mathematically analogous to a reaction diffusion
model for the reaction $A+B\to 0$ that was developed by the physicist John
Cardy. While this model is highly unrealistic, with a few modifications it
produces some qualitative features, such as heavy tailed price
distributions, that resemble their counterparts in real markets. 
\begin{figure}[ht]
\begin{center}
\includegraphics[scale=0.5]{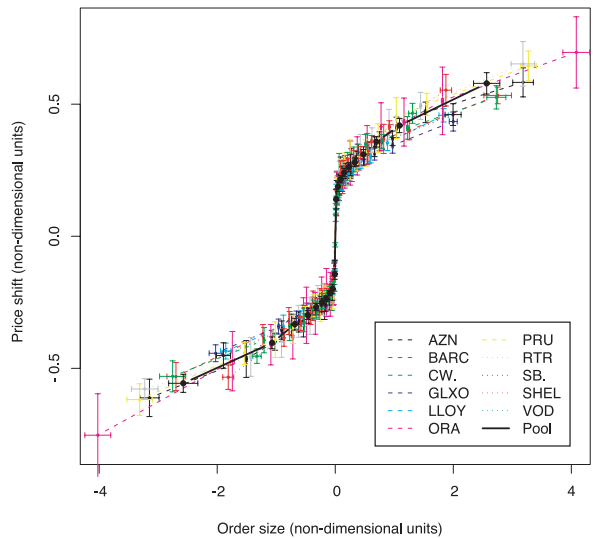}
\end{center}
\caption{ Nondimensionalized market impact as a function of trading size for
eleven stocks from the London Stock Exchange. The market impact is defined
as the price shift induced immediately upon arrival of a trading order
causing a transaction. Price is defined as the mean of the best quoted
prices to buy and to sell. Nondimensionalization of price and size is based
on a theory that treats arrival and cancellation of trading orders as a
stochastic queuing process, as described in ref.~\protect\cite{Farmer05}.
The pool is an average of the nondimensionalized data for all eleven stocks.
The collapse seen here indicates that after appropriate scaling all stocks
have a common price response. }
\label{fig:impact_collapse}
\end{figure}

Many variations of the BPS model have now been explored. One variation
simply assumes that agents place and cancel trading orders at random. After
performing a dimensional analysis based on prices, shares and time, the
resulting model can be shown to obey simple scaling laws that relate
statistical properties of trading order placement to statistical properties
of prices. These laws are restrictions on state variables similar to the
ideal gas law, but in this case the variables on one side are properties of
trading orders, such as the rates for order placement and cancellation, and
the variables on the other side are statistical properties of prices, such
as the diffusion rate in Bachelier's random walk model. These scaling laws
have been tested on data from the London Stock Exchange and have been shown
to be in surprisingly good agreement with it~\cite{Farmer05}. The model also
gives insight into the shape of supply and demand curves, as shown in Fig~%
\ref{fig:impact_collapse}.

\section{The El Farol bar problem and the minority game}

Another alternative approach is to develop highly simplified models of
strategic interaction that do a better job of capturing the essence of the
collective behavior in a financial market. Brian Arthur's El Farol bar
problem provides an alternative to conventional game theory. The name El
Farol comes from a bar in Santa Fe that is often crowded. Each day agents
decide whether or not to go hear music; if there is room in the bar they are
happy, and if it is too crowded they are disappointed. By definition only a
minority of the people can be happy, which leads to a phenomenon analogous
to \textit{frustration} - some desires are necessarily unsatisfiable and as
a result an astronomically large number of equilibria can emerge. The El
Farol model was simplified and abstracted by Challet and Zhang as the
minority game~\cite{Challet05}, in which an odd number $N$ of agents
repeatedly choose between two alternatives, which can be labeled $0$ or $1$.
Their decisions are made independently and simultaneously. Agents whose
choice is the minority value are rewarded (awarded ``payoffs'', in
game-theoretic terminology). Typically agents are capable of remembering the
outcomes of $M$ prior rounds of play, and maintain an inventory of $s$
strategies (random lookup tables) dictating a next move for each history.
For example, for $M=2$ a possible lookup table would be $00 \to 1$, $01 \to 0
$, $10 \to 1$, $11 \to 1$, meaning in the first case, ``If the previous
majority choices were $0$ in both previous rounds of the game, choose $1$ on
the next round". The strategy chosen is that with the best cumulative
performance.

\begin{figure}[ht]
\begin{center}
\includegraphics[scale=0.4]{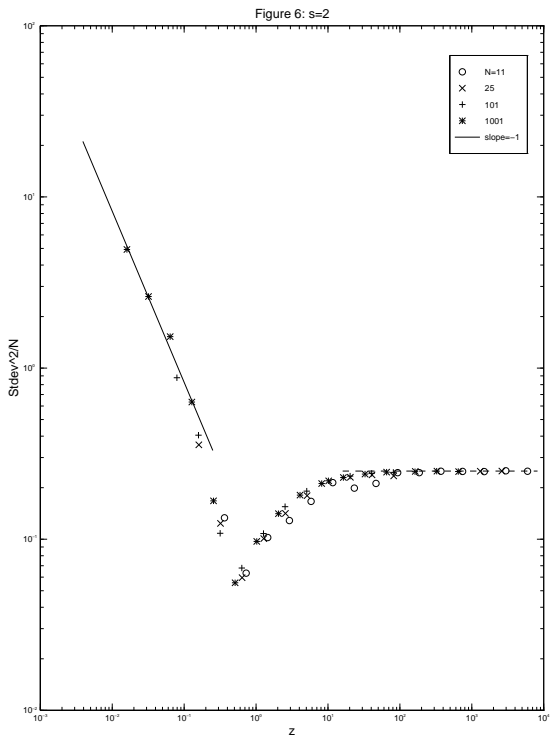}
\end{center}
\caption{ The variance of the number of winners in the minority game is
plotted against an order parameter $z = 2^M/N$. For $z < z_c$ the variance
is high, indicating that the game is highly inefficient in the sense that
the minority group is much smaller than it needs to be. The variance reaches
a minimum at $z = z_c \approx 0.4$, and when $z$ is large it approaches the
value it would have if all agents make random choices (from Savit et al. -
see reference \protect\cite{Challet05}). }
\label{fig:MG_phases}
\end{figure}

Minority games exhibit phase transitions for $s \ge 2$, in the ratio $z =
2^M/N$, of the number of resolvable pasts to the number of agents, as shown
in Fig.~\ref{fig:MG_phases}. 
For $z < z_c$ a critical value, the population is in a symmetric phase,
where the outcome of the next move is unpredictable from the history of
play. For $z > z_c$, in contrast, $N$ agents sparsely sample the space of
strategies, the next outcome is predictable, and the population is in a
symmetry-broken phase that can be understood analytically with replica
methods. The variance in the number of winners about the optimum, $\left(
N-1 \right) / 2$, measures the failure of ``allocative efficiency'', and is
minimized at $z_c$.

The minority game is readily extended to incorporate more features of real
financial markets, such as payoffs that increase as the size of the winning
group gets smaller (much as buyers or sellers of stocks can reap larger
profits when they are providing the more scarce of supply or demand), or the
``grand canonical'' version in which players are permitted to enter and
leave. With these enhancements the game self-organizes around the critical
point $z_c$, the payoff series exhibits fluctuations that display clustered
volatility, and they have a distribution with a power law tail, reminiscent
of a real market. The minority game provides a fascinating example of how a
very simple game can display a rich set of properties as soon as one moves
away from the rational choice paradigm.

\section{Entropy methods}

Finance is not the only area of economics where physicists are active. In
economics as in physics it is traditional to distinguish open from closed
systems, which give rise to different notions of equilibrium. Markets
considered merely as conduits for goods produced or consumed elsewhere are
described with theories of ``partial equilibrium'', largely specified by
open-system boundary conditions. Financial markets are open in this sense.
Economists also try to determine the ``general equilibria'' of whole
societies, taking into account not only trade, but production, consumption,
and to some extent regulation by government.

The understanding of relaxation to equilibrium, including when equilibria
are possible and whether they are unique, has grown in economics and in
physics together. In both fields mechanical models were used first, followed
by statistical explanations~\cite{Mirowski89}. Some recent work~\cite
{SmithFoley} has shown which subset of economic decision problems have an
identical structure to that of classical thermodynamics, including the
emergence of a phenomenological principle equivalent to entropy
maximization, while the more general equilibration problems usually
considered by economists correspond to physical problems with many
equilibria, such as granular, glassy, or hysteretic relaxation. The idea that equilibria correspond to statistically most-probable sets of configurations has led to attempts to define price formation
in statistical terms.  A related observation, that income distribution
seems consistent with various forms of entropy maximization, recasts
the problem of understanding income inequality, and interpreting how
much it really tells about the social forces affecting incomes (see
box~\ref{box:income_dist}).

We expect that maximum-ignorance principles will grow into a conceptual
foundation in economics as they have in physics, and that with this change,
the roles of symmetry, conservation laws, and scaling will become
increasingly important~\cite{Shubik_Smith}. Efforts to explain which aspects
of market function or regulatory structure converge on predictable forms,
relatively free of historical contingency, are likely to require
characterization in these more basic terms.

\section{Future directions}

Within the next few years we expect that in some physics and economics
departments a basic course teaching the essential elements of both physics
and economics will be designed (much as in biophysics; see Physics Today
March 2005). We believe physics will continue to contribute to economics in
a variety of different directions, ranging from macroeconomics to market
microstructure, and that such work will have increasing implications for
economic policy making.

One area of opportunity, where the applicability of physics might not be at
all obvious a priori, concerns the construction of economic indices, such as
the Consumer Price Index or the Dow Jones Industrial Average. Though these
indices provide only crude one dimensional summaries of very complex
phenomena, they play an important role in economic decision making. For
example, pension and wage payments are referenced to the CPI. Such indices
are currently constructed using essentially \textit{ad hoc} methods. We
believe that the accuracy of such indices could be improved by careful
thinking in terms of dimensional analysis, combined with better data
analysis correlating prices and other factors to the phenomena, such as
wages and pensions, for which the indices are designed. This is ultimately
related to the question of why the economy exhibits so many scale free
behaviors, such as the distribution of wealth or the size of firms. To shed
light on this we need a better understanding of the natural dimensions of
economic life, and the use of systematic dimensional analysis is likely to
be very useful in revealing this. Dimensional and scaling methods were a
cornerstone in the understanding of complex phenomena like turbulence in
fluids, and all the constituents that make fluid flow complex -- long time
correlations, nonlinearity, and chaos -- are likely to be even greater
factors in the economy.

At the other end of the spectrum, ideas from statistical mechanics could
make practical contributions to problems in market microstructure. For
market design, for example, some physics-style models suggest that changing
the rules to create incentives for patient trading orders vs. those that
demand immediate transactions could lower the volatility of prices. A
related practical problem concerns the optimal strategy for market makers,
i.e. agents that simultaneously buy and sell, and make a profit by taking
the difference. Though markets are increasingly electronic, the design of
automated market makers is still done in a more or less ad hoc manner. The
opportunity is ripe to create a theory for market making based on methods
from statistical mechanics. This could result in lowering transaction costs
and generally making markets more efficient.

We are reminded that several key ideas in physics are actually of economic
origin. A prejudice that the books should balance was likely responsible for
Joule's accounting for the energy content of heat before it was
well-supported by data. The concept of a ``currency'', which we still think
of primarily in economic metaphor, guides our understanding of the role of
energy in complex systems and particularly in biochemistry. Understanding
the dynamics and statistical mechanics of agency promises similarly to
expand the conceptual scope of physics.

\appendix

\section{Income distributions (BOX)}

\label{box:income_dist}

\begin{figure}[ht]
\begin{center}
\includegraphics[scale=0.45]{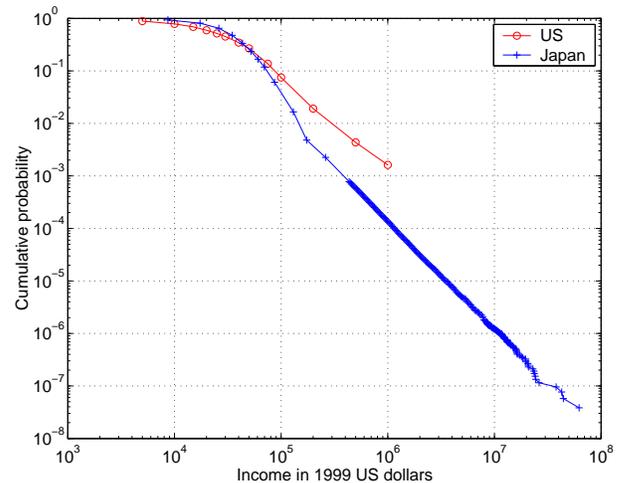}
\end{center}
\caption{ Modern examples of the Pareto distribution of incomes, from Ref.~%
\protect\cite{incomes}. Data are collected from federal income tax reporting
sources, and are more complete for high incomes in Japan than in the U.~S. A
systematic difference in recent data is that Japanese low incomes are better
fit by lognormal distribution, while U.~S.~low incomes are better fit by
exponential. }
\label{fig:Pareto_incomes}
\end{figure}

The first identification of a power law distribution -- in any field -- was
made by Vilfredo Pareto in 1897 for the distribution of income among the
highest earning few percent of inhabitants of the UK, and all income
distributions asymptotically of this form are known in economics as ``Pareto
distributions''. Subsequent studies by Pareto for Prussia, Saxony, Paris,
and few Italian cities confirmed these results, which continue to hold up
very well (see Fig.~\ref{fig:Pareto_incomes}). More recent studies~\cite
{incomes} have shown that not only is the income of the wealthy regular, but
so is the income of the majority of wage earners, and the two groups follow
different distributions. The low- and medium-income body of the distribution
is either exponential or lognormal (variable among data sets), with a
transition to the Pareto law for large incomes, at a level that varies with
time, tax laws, and other factors, as yet unknown. 

As striking as the fact that the large-income distribution is scale free, is
the fact that the overall distribution is so featureless, being described by
four (or five) parameters: mean income, Pareto exponent, transition point
between low- and high-income ranges, and the exponential constant (or mean
and variance of the lognormal) in the low range. Pareto, lognormal, and
exponential distributions are all limiting distributions of simple random
processes, and can also be derived as maximum-entropy distributions for
either income or its logarithm, subject to appropriate boundary conditions
on the (arithmetic or geometric) mean income~\cite{Yakov}.

Income distribution is a hot topic economically and politically, because it
lies at the heart of a society's notions of egalitarianism, opportunity, or
social insurance. Not surprisingly, causes of income inequality are
asserted, such as distinctions between capital ownership and wage labor,
with major policy implications. Maximum-entropy interpretations of income
distribution place conceptual as well as quantitative bounds on these
arguments. They suggest that the many detailed features of a society that
could in principle affect incomes somehow average so that their individual
characteristic scales are not imprinted on the aggregate distribution; the
ultimate constraints may be conservation laws or boundary conditions
reflected in at most a few parameters. Such featureless averaging, like the
scaling relations we have noted above, may suggest that a form of
universality classification is fundamental to understanding economics, as it
has been to thermodynamics and field theory.

\section{Option pricing (BOX)}

\label{box:option_pricing}

Bachelier's random walk was a triumph of quantitative finance, and became
the basis of modern portfolio analysis, and later the Black-Scholes model
for option pricing~\cite{Bouchaud00}. The $\alpha $ and $\beta $
coefficients published in every security analysis are mean and covariance
coefficients from fits to a random walk. However, the heavy tails of real
price fluctuations, under-predicted by the Gaussian distribution resulting
from accumulation of an uncorrelated random walk, can lead to disastrous
mis-estimates of risk. This has been an important problem in financial
mathematics which has received a great deal of attention.

\begin{figure}[ht]
\begin{center}
\includegraphics[scale=0.5]{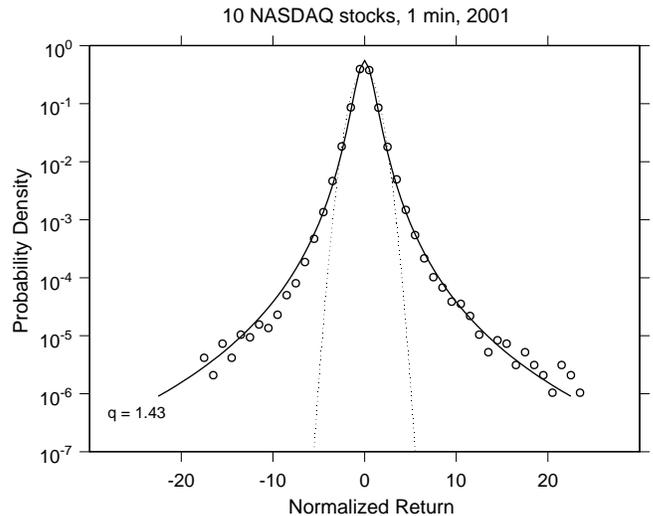}
\end{center}
\caption{ Distributions of log returns for 10 Nasdaq high-volume stocks.
Returns are calculated as $\log p(t + \tau) - \log p(t)$, where $p$ is the
transaction price and $\tau$ is one minute, and are normalized by the sample
standard deviation. Also shown is the student distribution (solid line)
which provides a good fit to the data (from ref.~\protect\cite{Borland02}). }
\label{fig:student_returns}
\end{figure}

\begin{figure}[ht]
\begin{center}
\includegraphics[scale=0.5]{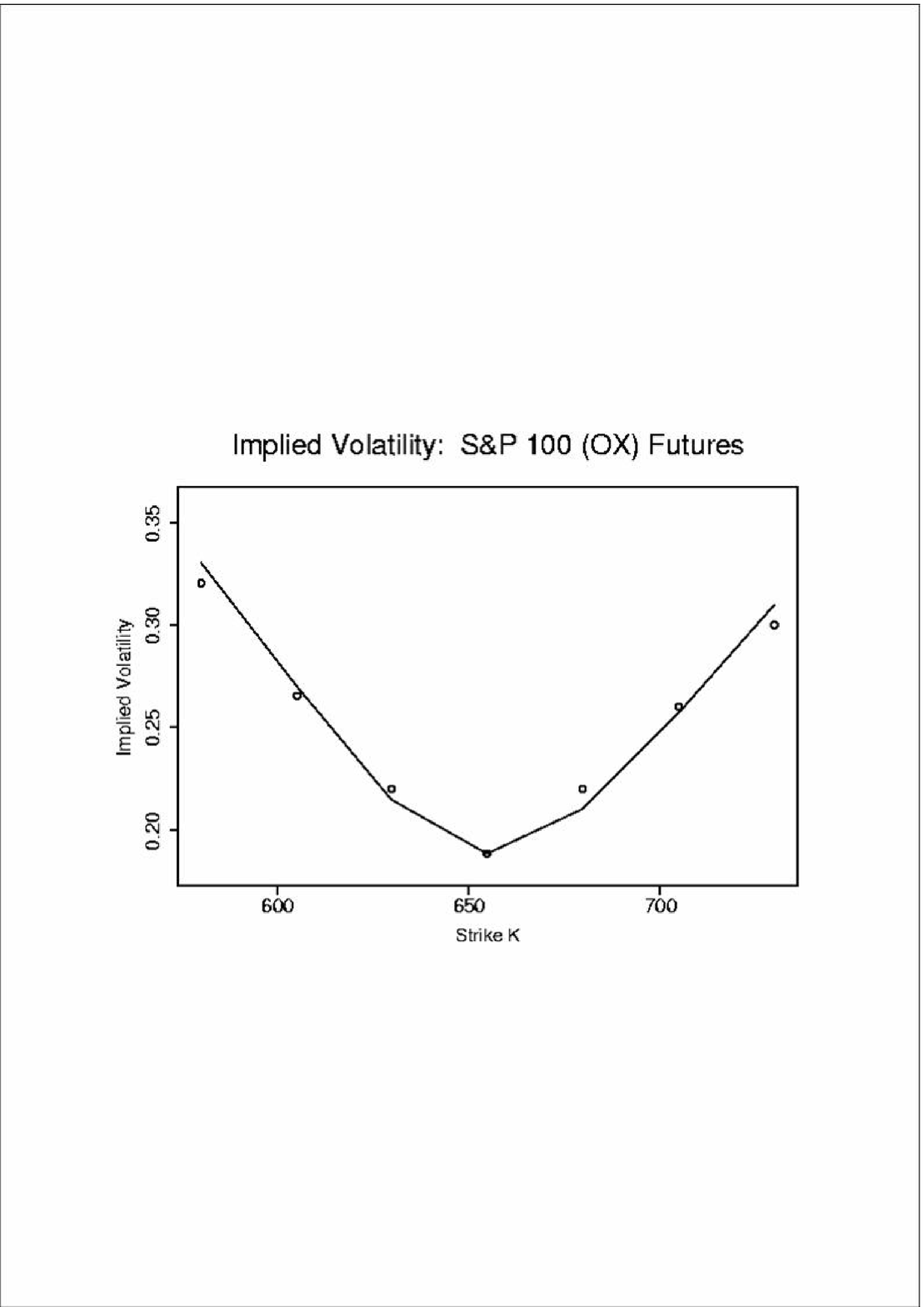}
\end{center}
\caption{ A typical example of the "smile" of the implied volatility (sigma)
needed in the Black-Scholes pricing formula to correctly price options,
versus the strike price. The example shown here corresponds to that of June
2001 call options on S\&P 100 futures (OX), with 10 days to expiration
(symbols). This is well-fit by the theory developed by Borland (line). }
\label{fig:vol_smile}
\end{figure}

More recent work by physicists extends analytic methods for pricing options
to take the heavy tails and volatility bursts of real prices into account.
Inspired by the work of Constantino Tsallis on non-extensive statistical
mechanics~\cite{Gell-Mann04}, (see Fig.~\ref{fig:student_returns}) Lisa
Borland \cite{Borland02} has developed a new pricing formula that corrects
the standard Black-Scholes model. For three decades option-pricing
practitioners have recognized that random-walk estimates were too
conservative, and compensated by altering the parameters of the
Black-Scholes model depending on the strike price of the option. The
``implied volatility'' assigned in this way makes a well-known ``smile''
when plotted versus the strike prices (see Fig.~\ref{fig:vol_smile}). The
Borland option-pricing formula provides a rational basis for pricing of rare
events, and nicely reproduces the volatility smile. While there are a large
number of other generalizations of the Black-Scholes theory that address the
problem of the smile, Borland's has the significant advantage that it gives
a closed form solution.

The self-consistency condition on which all rational option pricing is based
-- arbitrage-free hedging of risk -- is a classically reductionist principle
relating derivatives to their underlying assets. It is noteworthy that
economic practice has chosen to adhere to the specific Black-Scholes formula
based on an empirically invalid model of the underlying, and to introduce
the phenomenological curve of ``implied volatility'' to bring the formula in
line with data. The Borland construction avoids such mixing of reductionism
and phenomenology; its single additional parameter describes the observed
fluctuations in the underlying asset price, separating the problem of
explaining these from that of pricing their derivatives.


\end{document}